\documentclass[11pt,notitlepage]{article}
\newcommand{\preprint}{} 

\usepackage{akshay}
\usepackage{float, caption, subcaption}
\usepackage{algorithm, algorithmic}

\ifdefined \preprint
	\usepackage[left=1.25in,right=1in,top=1in,bottom=1in]{geometry}
	\usepackage[title,header]{appendix}
	\usepackage{akshayoptional}
	\usepackage{setspace}
\else\ifdefined \informs
\else
\fi
\fi

\newcommand{\R}{W}
\newcommand{\utilitysym}{\phi}
\newcommandx{\utility}[2][1={ij},2=p_j,usedefault]{\utilitysym_{#1}(#2)}
\newcommand{\utilityf}[1][i]{\utilitysym_{#1}}
\newcommand{\utilSing}[1][i]{\varphi_{#1}}

\newcommand{\newphi}{\tilde{\utilitysym}}
\newcommandx{\valuation}[2][1=i,2=j,usedefault]{\nu_{#1 #2}}
\newcommandx{\valuef}[2][1=,2=i]{
\ifthenelse{\isempty{#1}}
{\nu_{#2}}
{\nu_{#2}(#1)}
}

\newcommand{\itemcard}{s}

\newcommand{\J}{M}

\newcommand{\level}{\ell}

\newcommand{\A}{\mathcal{A}}
\newcommand{\M}{\mathcal{M}}
\newcommand{\D}{\mathcal{D}}

\newcommand{\B}{\mathcal{B}}

\newcommand{\pl}{p^{L}}

\newcommand{\m}{N}
\newcommand{\n}{M}

\setvalue{\MyTitle}{Optimal prices for profit-maximizing envy-free allocations with general utilities}
\setvalue{\MyAbstract}{
We consider the envy-free pricing problem that arises in the setting of economic equilibrium of single-seller, multi-item, multi-buyer markets. With linear utility functions, this problem has been extensively studied in literature for computational complexity and approximation algorithms. We work within the framework of general utilities and derive many properties of optimal prices and extend some complexity results. Our main theoretical results are with all buyers having unit demand. We establish a chain property on the optimal prices when supply is unlimited, pseudo-polynomial solvability when there is a certain monotonicity condition on utility functions, NP-hardness when each buyer desires at least two items, and polynomial solvability when each buyer is single-minded and supply is unlimited.
}
\setvalue{\MyKeywords}{Envy-free, Fair division, Nash equilibrium}
\setvalue{\MySubjclass}{}
\setvalue{\MyDate}{Manuscript draft: 25 November 2026}

\setvalue{\NumAuthors}{2}
\renewcommand{\authordef}[1]{%
\ifcase#1\relax \setvalue{\authorname}{Akshay Gupte} \setvalue{\authoraffil}{School of Mathematics and Maxwell Institute for Mathematical Sciences, University of Edinburgh, UK} \setvalue{\authoremail}{agupte@ed.ac.uk}

\or \setvalue{\authorname}{Yiran Zhu}

\else Jane Doe
\fi
}

\begin{document}

\ifdefined \preprint
	\author{\authordef{0}\MyAuthor \and \authordef{1}\MyAuthor[1]}
	\title{\MyTitle}
	\date{\MyDate}
	\maketitle

	\begin{abstract}
	{\MyAbstract}
	\ifthenelse{\equal{\MyKeywords}{}}{}{\keywrds{\MyKeywords}}
	\ifthenelse{\equal{\MySubjclass}{}}{}{\amsclass{\MySubjclass}}
	\end{abstract}
\fi


\section{Introduction}

We consider economic equilibrium problems arising in single-seller, multi-item, multi-buyer markets. 
The setting is as follows. A seller has many items to sell to multiple 
buyers (also called agents). Each buyer is interested in buying at most one item and has a known valuation for each item (generally construed as the maximum price that the buyer is willing to pay for that item). Multiple buyers can buy the same item because there are multiple copies of each item. The goal is to decide allocations of items to buyers and the prices that the seller must charge for each item with the objective of maximising their revenue by selling as many items as they can. A buyer will not buy an item if their utility value for that item is negative, where in the simplest form this means that their valuation must exceed the item price set by the seller. For an allocation to be feasible, we also need some form of fairness and equilibrium amongst the buyers so that they do not have an incentive to switch to a different allocation. 


A classical approach in economics for solving this problem  has been the notion of \emph{Walrasian equilibrium} \cite{walras1874elements}, also called competitive equilibrium, which is to find allocation and prices such that all buyers are \emph{envy-free} and the market clears (unmatched items are priced at zero). A buyer is envy-free if their utility cannot be increased by switching to another item. Here, utility of buyer $i$ for item $j$ is measured by a utility functional 
for which a typical choice in literature is that of \emph{linear utilities} which means the utility of a buyer for an item is the difference of their item valuation with the selling price of the item. 
Walrasian equilibrium achieves fairness for each buyer through its envy-free requirement and also guarantees efficiency for the entire market through its market clearing condition. However, it may not exist in general unless special assumptions are made \cite{nisan2007algorithmic,shapley1971assignment}, and even if it exists, the total revenue, which is calculated from the seller's perspective as the total price of matched items (counting multiplicity of matching), generated by it can be very low. To counter this, another equilibrium notion called \emph{envy-free pricing} was proposed by \citet{guruswami2005profit}. This relaxes the market clearing condition and imposes only envy-freeness of buyers, thus guaranteeing that an envy-free equilibrium pricing always exists, while maximizing the total revenue. The envy-free pricing model of \citeauthor{guruswami2005profit} is closely related to the problem of bundle pricing commonly found in economics and management science \cite{adams1976commodity,mcafee1989multiproduct,hanson1990optimal}. The idea is that when selling different items, the seller may choose to adopt a mixed bundling strategy, which is to say that he/she may sell a family of subsets of items, with different pricing on different subsets. This is opposed to pure bundling, which offers a fixed bundle at a set price that encompasses all the goods. The buyer can then either decide to buy the bundle and get all the included items in it or to not buy the bundle and get none of the items. The seller prices the items from which bundles are created and typically, the additive law is followed for pricing each bundle, i.e., it is set equal to the sum of prices of individual items in it. If the seller prices $n$ items, then the mixed bundling problem presents at most $2^{n}$ choices for each buyer. One way to keep the number of choices polynomial in input is to use \emph{cardinality bundling} \cite{hitt2005bundling,wu2008customized,chu2011bundle}, which is cardinality-agnostic since subsets of same cardinality are assigned the same price regardless of which items are in the subset. 

The envy-free pricing problem has garnered a lot of attention in terms of hardness results and approximation algorithms. The problem is NP-hard and APX-hard in general, and remains so even under various assumptions on valuations, such as each buyer having positive valuation for a fixed number of items  \cite{balcan2006approximation,chen2014envy,cheung2008approximation,guruswami2005profit,chen2011metric}. Conditions have been established for the existence of envy-free allocations \cite{manurangsi2020envy}. The other focus area has been the special case of cardinality bundling (which has linear utilities), and here a very weak Lagrangian-based relaxation \cite{wu2008customized}  and some simple integer programming-based heuristics \cite{Dobson93heuristics} have been implemented. A special case of cardinality bundling that has the matrix of valuations $\valuation$ satisfying the Monge property was analysed recently \cite{wu2019cardinality} and it was shown that the problem can be solved in polynomial time in this case.

\subsection{Our Results.}
\textsection\ref{sec:setup} presents a generalised approach to studying envy-free problems without assuming linear utility functions. In addition to the equilibrium condition, we also impose demand sets on each buyer which can include maximum budget for purchasing items, cardinality constraint on the number of items that can be bought, etc.
\textsection\ref{sec:prices} and \textsection\ref{sec:optimal} establish properties of envy-free prices including a utility-chain structure (\Cref{utilitychain}) that is a generalisation of the linear case from literature, and an integrality property (\Cref{intPrice}). Our complexity analysis is in \textsection\ref{sec:complexity} and summarised in \Cref{tab:complexity} in context of existing literature. In particular, under different problem conditions, we have a NP-hardness result in \Cref{thm:np-hard}, polynomial solvability in \Cref{thm:poly-solvable}, and pseudo-polynomial solvability in \Cref{thm:pseudo-poly}.

\section{Problem Setup}	\label{sec:setup}

There are $n\ge 1$ buyers choosing from $m\ge 1$ items offered by a single seller. 
These are indexed by $\m := \{1,\dots,n\}$ and $\n := \{1,\dots,m\}$. The item set is partitioned into $r\ge 1$ categories as 
$\n = \bigcup_{k=1}^{r}\J_{k}$ with $\J_{i}\cap \J_{k}=\emptyset$, where each $\J_{k}$ corresponds to a collection of items of the same type. For each item $j\in\n$, there is a fixed cost $d_{j} \ge 0$ for selling at least one copy of $j$, a variable cost $c_{j} \ge 0$  that is incurred every time a copy of $j$ is sold to some buyer, and at most $\itemcard_j \ge 1$ copies of item $j$ can be sold. The seller has to determine the selling price (price per copy)  of each item in order to maximise their revenue. Prices are assumed to be additive in the sense that the total price for a subset of items is equal to the sum of price of each item in the subset. Buyers have to decide which items they wish to buy and can buy at most one copy of an item since they do not gain any additional value from purchasing more copies. An item that is not bought by anybody is not sold. Every buyer $i$ has a demand set $\D_{i} \subseteq 2^{m}$ which is the collection of allocations that the buyer considers to be suitable. This demand set could depend on the prices set by the seller, for e.g., when it represents budget constraints which restrict how much money the buyer can spend, or it could be independent of the prices.

There are two kinds of decision variables, item prices and allocations of items to buyers. A feasible price is any nonnegative vector $p\in\real^{m}_{+}$. The additivity assumption on prices implies that the total price of a subset is \[p_{S} := \sum_{j\in S}p_{j}, \qquad S\subseteq\n.\] We denote a market allocation by the tuple $\A := (A_{1},A_{2},\ldots,A_{n})$ where $A_{i}\subseteq\n$ are the items that buyer $i$ has decided to buy. For convenience, we partition $i$'s allocation into the different product categories by denoting
\begin{subequations}	\label{def:AM}
\begin{equation}	\label{def:A}
A_{i} = \left(A_{i1}, A_{i2}, \ldots, A_{ir} \right), \quad \text{where } \ A_{ik} := A_{i}\cap \J_{k}, \ \; k\in[r] .
\end{equation}
The buyers who have bought item $j$ are denoted by 
\begin{equation}
B_{j} = B_{j}(\A) := \left\{i\in\m : j \in A_{i} \right\}, \quad j\in\n . 
\end{equation}
The set of all buyers who have bought any item is denoted by $\B = \cup_{j\in \n} B_j$.
The matched items, i.e. the items that are bought by at least one buyer, are
\begin{equation}	\label{def:M}
\M = \M(\A) := \bigcup_{i\in\m}A_{i} = \left\{j\in\n : B_{j} \neq \emptyset \right\} .
\end{equation}
The matched items in category $k$ is denoted by
\begin{equation}
    \M_k(\A):= \M(\A) \cap \J_k, \quad k\in [r]
\end{equation}
\end{subequations}
These are also the items the seller has decided to sell. There is no requirement for the market to clear and hence some items can remain unsold ($\M$ can be a strict subset of $\n$) and a buyer can go unallocated ($A_{i}$ can be empty).

A feasible market allocation requires that an item not be oversold beyond the supply of number of copies, and $A_{i} \in \D_{i}$ for all $i$. Common examples of $\D_{i} \subsetneq 2^{m}$ are cardinality constraints that restrict the number of items that the buyer can buy, such as $\D_{i} = \{S\subseteq\n : |S| \le b_{i}\}$ or $\D_{i} = \{S\subseteq\n : |S\cap \J_{k}| \le b_{ik}, \; k\in[r] \}$, or budget constraints on how much money the buyer can spend $\D_{i} = \{S\subseteq\n : \sum_{j\in S}p_{j} \le b_{i}\}$ (this knapsack constraint also restricts them from buying too many items).  

The main challenging ingredient of our problem is envy-freeness for fair division amongst buyers. We will make this explicit in the next few paragraphs using utility functions. We also impose rational thinking of the buyers.

Every buyer $i\in\m$ has a valuation function $\valuef:2^{\n}\to \real_{+}$ 
where $\valuef[S]$, for $S\subseteq\n$, denotes the value to $i$ for choosing all the items in $S$. This can also be regarded as the maximum price they are willing to pay for that subset. We don't assume any particular structure on these valuations, such as submodularity, etc. The possibility of a buyer having valuations for only some specific subsets can easily be accommodated by setting those valuations equal to zero. No value is obtained for not choosing any items and so $\valuef[\emptyset] = 0$.

There is also a utility function $\utilityf: 2^{\n} \times \real^{m}_{+}\to\real_{+}$ where $\utilityf(S,p)$ is the utility to buyer $i$ for choosing all the items in $S\subseteq\n$ when the seller has priced the items with $p\in\real^{m}_{+}$. Since the item price is determined by the seller and is the decision variable in our problem, the utility values are also unknown quantities in our problem. We assume that this utility can be calculated in terms of the valuation function and the total price $p_{S} := \sum_{j\in S}p_{j}$ of the items in the chosen subset $S$.

\begin{assumption}[Modelling utilities]	\label{utilassume}
For each $i\in\m$, there exists a non-decreasing univariate function $g_{i}:\real_{+}\to\real$ and a bivariate function $f_{i}:\real^{2}\to\real$ so that we can write the utility values as
\[
\utilityf(S,p) = \utilSing(S, p_{S}): = f_{i}\left(\valuef[S], g_{i}(p_{S}) \right).
\]
Each utility function is non-decreasing in valuations and non-increasing in prices, which is achieved by having $f_{i}(\cdot,b)$ to be non-decreasing for every $b\in\real$ and $f_{i}(a,\cdot)$ to be non-increasing for every $a\in\real$. Furthermore, $f_{i}(0,g_{i}(0)) = 0$ so that there is zero utility for choosing nothing.
\end{assumption}

We are now ready to formally define the two challenging ingredients that impose constraints in our problem. The first requirement relates to rational-mindedness of a buyer, which allows a buyer to reject items whose utility is considered as being too low. 

\begin{definition}[Rationality in buying]	\label{def:rational}
Each buyer $i\in\m$ has a preferred level $\level_{ik} \ge 0$ for each product category $k \in [r]$ to indicate that they will choose only those items that give them at least this much utility at the prices set by the seller. Thus, for given item prices $p\in\real^{m}_{+}$, an allocation $A_{i}$ must satisfy the following for all $k\in[r]$,
\[
A_{ik}\neq\emptyset \quad \text{only if } \quad \utilityf(A_{ik},p) \;\ge\; \level_{ik} . 
\]
\end{definition}

Nonnegativity of $\level_{ik}$ means that items of negative utility value are always undesirable. To avoid items of zero utility being bought, a buyer can set $\level_{ik}$ to be strictly positive.

Our second requirement, which is the main emphasis of this work, is on creating allocations that result in each buyer being in an equilibrium with their selection. This is captured by the notion of \emph{envy-freeness} which we explain in the next two definitions.

\begin{definition}[Envy-free buyer]\label{def:envy-buyer}
For given prices $p\in\real^{m}_{+}$, items $J\subseteq\n$ that the seller has decided to sell, and allocation $A_{i}$ of buyer $i$, we say that the buyer is {envy-free in the $k^{th}$ product category}, for $k\in[r]$, if either of the following holds,
\begin{enumerate}
\item $A_{ik} = \emptyset$ and $\utilityf(S\cap\n_{k}\cap J,p) < \level_{ik}$ for all non-empty $S\in\D_{i}$, or \smallskip
\item $\utilityf(A_{ik},p) \ge \utilityf(S\cap\n_{k}\cap J,p)$ for all non-empty $S\in\D_{i}$. \smallskip
\end{enumerate}
Then, buyer $i$ is said to be \emph{partially envy-free} if they are envy-free for some category, and they are said to be \emph{totally envy-free} if they are envy-free for all categories.
\end{definition}

\begin{definition}[Envy-freeness]	\label{def:envy}
Prices $p\in\real^{m}_{+}$ are \emph{envy-free} if there is a feasible market allocation such that all buyers are \emph{totally envy-free}.
\end{definition}



\subsection*{Nomenclature}

There are many types of utilities that satisfy \Cref{utilassume}. Literature on this topic commonly uses \emph{linear utility} functions which are defined as the valuation function being additive, each $g_{i}$ an identity function, and each $f_{i}$ as the difference of its two arguments, so that we have
\begin{equation} \tag{Linear utility}
\utilityf(S,p) = \sum_{j\in S}\valuef[\{j\}] - \sum_{j\in S}p_{j} = \sum_{j\in S}\valuation - p_{j},
\end{equation}
where we write $\valuef[\{j\}] := \valuation$ for convenience. tion that perhaps appears in many practical problems, but our analysis in this paper reveals that the linear utility assumption is not always needed. We can also handle more general utility functions  such as the difference of valuation and some function of item price, which subsumes the linear case. Considering utility as the surplus of valuation over price seems like a reasonable assumpHowever, we will need to make the mild assumption that the utility function can attain the lowest preferred level of the buyer. In particular, this means that if the buyer can accept an item giving them zero utility, which is the lowest-level that rational behaviour would warrant, then the utility function has a nonnegative root. 

\begin{assumption}[Attaining lowest utility]	\label{asmp:finite A}
For each $i\in \m, k\in [r], S\subseteq \J_{k}$, there exists a finite $\R_{i,S}\ge 0$ such that $f_i(j, \R_{i, S})=\level_{ik}$
\end{assumption}

Thus, the utility function like $\utilSing(S, p_{S}) = \valuef[S]/p_{S}$ is no longer valid when $\valuation > 0$ even though it satisfies the monotonicity condition. This precludes the possibility that customers may possess a positive utility for arbitrarily high prices. But separable utility functions satisfy both conditions.

\begin{definition}[Separable utilities]
A separable utility function is of the form $\newphi : (\nu,p)\in\real^{2}_{+}\to\nu - g(p)$ for some increasing function $g : \real_{+}\to\real$ and scalar $\nu$ that belongs to the range of $g$. 
\end{definition}

The root of a separable utility is given by $\R = g^{-1}(\nu)$. The linear case corresponds to taking $g$ as the identity map, for which the valuation is the root. 


\subsection*{Combinatorial formulation}

A feasible price is a nonnegative vector $p\in\real^{m}_{+}$ for which there exists a market allocation satisfying supply constraints on the items, demand set of each buyer, rational-mindedness as per \Cref{def:rational} and envy-freeness as per \Cref{def:envy}. We refer to a feasible solution by the tuple $(p,\A)$. The combinatorial optimization formulation can be described as follows.  
\begin{subequations}\label{eq:prob}
\begin{align}
\max_{p,\A} &\quad \sum_{j\in\M}\, \left[ \left(p_{j} - c_{j}\right)|B_{j}| \, - \, d_{j}  \right] \label{eq:prob:obj} \smallskip\\
\text{s.t.} &\quad \A, B \text{ and } \M \text{ as per } \cref{def:AM} \label{eq:prob:set-def} \\
&\quad (p, \A) \text{ are envy-free as per } \Cref{def:envy} \label{eq:prob:envy} \\
&\quad \utilityf(A_{ik},p) \ge \level_{ik}, \qquad i\in\m, k \in [r] \text{ s.t. } A_{ik}\neq\emptyset \label{eq:prob:min-level}\\
&\quad A_{i}\in \D_{i}, \qquad i\in\m \label{eq:prob:cardi}\\
&\quad \left|B_{j} \right| \le \itemcard_j, \qquad j\in\n. \label{eq:prob:supply}
\end{align}
\end{subequations}
The objective is the total revenue from selling minus the cost of selling. The first constraint enforces the envy-free equilibrium condition where each buyer is allocated an item for which they have maximum utility, whereas the second constraint enforces the rationality of a buyer who will not buy an item that provides negative utility. The last two constraints are buyer's budget and item supply.

We also analyse the combinatorial structure of the classic \textit{unit-demand, unlimited supply} envy-free problem studied in literature \cite{wu2008customized, guruswami2005profit}. Here, \textit{unit-demand} means the buyer can buy at most 1 element from each category $M_{k}$ and \textit{unlimited supply} means the seller can sell multiple copies of the same item to different buyers.
    \begin{align} \label{eq:prob-ep}
    \max \;\cref{eq:prob:obj}, \quad \cref{eq:prob:set-def} - \cref{eq:prob:supply}, 
    \quad \D_{i} = \left\{S\subseteq\n : |S\cap M_k| \le 1 \right\} .
    \end{align}

\section{Feasible Envy-free Prices}	\label{sec:prices}

We being with the following straightforward observation on the prices.
 
\begin{lemma}	\label{unmatchprice}
When the prices  are fixed for all matched items, then increasing the prices for non-matched items does not alter the solution or the objective value.
\end{lemma}
\begin{proof}
Non-matched items are characterised by having a utility lower than the preference level or by offering the maximum utility of other items within the same category. An increase in the prices of non-matched items further reduces their utility, making them even less attractive to buyers. Since these items are not selected under current pricing, elevating their prices ensures they remain unchosen, leaving the original solution and objective value unaffected.
\end{proof}


When all the buyers are assumed to be rational as per \Cref{def:rational}, the condition of envy-free pricing reads as follows.

\begin{lemma} \label{lem:envy}
Prices $p\in\real^{m}_{+}$ are {envy-free} in a market with rational-minded buyers if and only if there is a feasible market allocation $\A$ such that 
\begin{equation}	\label{eq:envy}	
\max\left\{\level_{ik} -\varepsilon, \, \utilityf(A_{ik},p)\right\} \;\ge\; \utilityf(S,p), \qquad i\in\m, \; k\in [r], \; S\subseteq\D_{i}\cap\M_{k}(\A) .
\end{equation}
for some $\varepsilon >0$.
\end{lemma}
\begin{proof}
    If $p$ is envy-free, then the claimed inequality holds for the case whenever $\max\{\utilityf(S,p): S\subseteq \D_i\cap \M_k(A)\} \ge \level_{ik}$ because its l.h.s. is greater than the r.h.s. in \Cref{def:envy}. In the case that $\level_{ik} > \utilityf(S,p)$ for all $S\in \D_i\cap \M_k(\A)$. Let $\varepsilon_{ik} = \min\{\level_{ik}-\utilityf(S, p): S\subseteq \D_i\cap \M_k(\A)\}$. We then define $\varepsilon = \min\{\varepsilon_{ik}: i\in \m, k\in [r]\} > 0$. 
    Then l.h.s. in the claimed inequality is at least $\level_{ik}-\varepsilon$ which is larger than $\utilityf(S, p)$ by construction of $\varepsilon$.  
    
    Now let the inequality hold. To argue that $p$ is envy-free we have to show that the feasible market allocation $\A$ satisfies conditions in \Cref{def:envy}. For any $i\in \m, k\in [r]$, consider $\max\{\utilityf(S,p): S\subseteq \D_i\cap \M_k(\A)\} < \level_{ik}$. From the rationality of buyers, $A_{ik}=\emptyset$ since $\utilityf(A_{ik}, p) < \level_{ik}$. Observe that \cref{eq:envy} should hold for any $0 <\varepsilon' < \varepsilon$ and so, we can choose a sufficiently small $\varepsilon' < \level_{ik} - \max\{\utilityf(S,p): S\subseteq \D_i\cap \M_k(\A)\}$. 
    then $\max\left\{\level_{ik}-\varepsilon',\, \utilityf(A_{ik},p)\right\} = \level_{ik}-\varepsilon'$ since $\utilityf(A_{ik}, p) \le \max\{\utilityf(S,p): S\subseteq \D_i\cap \M_k(\A)\}< \level_{ik} - \varepsilon'$. Then \cref{eq:envy} implies that $\level_{ik}> \utilityf(S, p)$ as required in \Cref{def:envy} for empty allocation. For the case  $\max\{\utilityf(S,p): S\subseteq \D_i\cap \M_k(\A)\}\ge \level_{ik}$. Then we have
    \[
    \max\left\{\level_{ik} -\varepsilon, \, \utilityf(A_{ik},p)\right\} \ge \max\{\utilityf(S,p): S\subseteq \D_i\cap \M_k(\A)\} \ge \max\{\utilityf(A_{ik}, p), \, \level_{ik}\}
    \]
    This implies $\max\{\utilityf(A_{ik},p), \level_{ik}-\varepsilon\} = \utilityf(A_{ik},p)$, and therefore $\utilityf(A_{ik}, p) \ge \utilityf(S, p)$ which is the same as in \Cref{def:envy}. Consequently, $p$ is envy-free.
\end{proof}

Next we derive an upper bound on the prices using the maximum willingness across all buyers.

\begin{lemma}\label{lem:boundprice}
A feasible envy-free price satisfies the following upper bound,
    \[
    p_{j} \le \max \left\{W_{i,S}: i\in \m, j\in S\subseteq \n_{k}\right\}, \qquad j\in \n
    \]
\end{lemma}
\begin{proof}
   Let $\R_{i, j}^* = \left\{\R_{i,S}: i\in \m, j\in S\subseteq \n_{k}\right\}$. We begin by establishing $p_{j}\le \min\{\R_{i,j}^{*}: i \in B_j \}$ for $j \in \n_{k}$ and so $p_{j}\le \R_{i,j}^{*}\le \max\{\R_{i,j}^*: i\in \n\} $. Recall that $A{ik}$ represents items in the $k$-th category allocated to buyer $i$. If $i\in B_j$, then $j\in A_{ik}$. The minimum level constraint \cref{eq:prob:min-level} is $\utilSing(A_{ik}, p_{A_{ik}}) \ge \level_{ik}$. Given $\utilSing(A_{ik}, \R_{i, A_{ik}})=\level_{ik}$ and the decreasing nature of utility with price, we have $p_{j}\le p_{A_{ik}} \le \R_{i, A_{ik}} \le \R_{i, j}^{*}$. For items not for sale ($B_{j}=\emptyset$), their price does not impact the objective value, thus the upper bound trivially holds.
\end{proof}


\section{Optimal Prices for Unit Demand, Unlimited Supply} \label{sec:optimal}


This section analyses properties of optimal prices for the problem stated in \cref{eq:prob-ep}. As a relaxed version of \cref{eq:prob}, this unit-demand and unlimited-supply problem afford more structure on its optimal envy-free prices. We show that its optimal pricing follows a chain equality in terms of utilities (see \Cref{utilitychain}). For the unit-supply market, the chain equality has been proved previously in the linear utility settings \cite{chen2014envy}. It is crucial in proving the polynomial-solvable for the market where each buyer prefers at most 2 items \cite{chen2014envy}. In this paper, we extend this result to the unlimited-supply setting. In particular, our results remains valid in the case even when considering non-linear utilities and listing costs.

%
%

\begin{proposition}\label{prop:priceLowerBound}
When there is unit-demand and unlimited-supply, every optimal price $p$ satisfies the lower bound $p_{j} \ge c_{j}$ for all $j$.
\end{proposition}

\begin{proof}
Let us denote $\pl := c$.  We prove by contradiction that the lower bound $\pl \le p$ should be true for all optimal solutions. Suppose there is an optimal solution $(\A, p)$ with optimal pricing $p$ such that $p_j < \pl_j$ for some $j\in \M$. In particular, we denote $J:=\{j\in \M: p_j < \pl_j\}$ for all such items with price lower than $\pl$. 
Let $I=\bigcup_{j\in J} B_j$ be the set of all buyers choosing items from $J$. w.l.o.g. we can assume $I\neq \emptyset$, otherwise we can adjust $p_j=\pl_j$ for all $j\in J$ while still maintaining optimal. 

The main idea is to withdraw these items in $J$ and re-allocate the items available to the buyers from $I$ subject to envy-free constraints \cref{eq:prob:envy} and minimum level requirement \cref{eq:prob:min-level}.  \Cref{unmatchprice} tells us that increasing the prices of not-for-selling items has no influence on the overall objective. Therefore, we define the new price $p'_{j} \coloneqq \max\{c_{j}, p_{j}\}$ for $j\in J$. The new allocation $\A'$ is defined by letting 
\begin{equation}\label{prop:priceLowerBound:eq:new-allo}
    A'_{ik} = \arg\max\{\utilityf(j, p'): j\in \M_{k}(\A)\setminus J) , \utilityf(j, p')\ge \level_{ik}\} 
\end{equation}
and we define the argmax of empty set being empty set. Note that $\utilityf(A'_{ik}, p') \ge \level_{ik}$ whenever $A'_{ik}\neq \emptyset$. To verify the new solution $(\A', p')$ is feasible, it is sufficient to verify the envy-free condition. We first prove that $\M(\A') = \M(\A)\setminus J$. From the construction of $\A'$, it's clear that $\A' \subseteq \M(\A)\setminus J$. To see the other direction, for each $j\in \M(\A)\setminus J$, there exists $k\in [r]$ such that $j\in \n_{k}$ and also $B_{j}\neq \emptyset$. For each $i\in B_{j}$, the feasibility of $(\A, p)$ implies that $\utilityf(j, p) \ge \utilityf(h, p)$ for all $h\in \M_{k}(\A)$ and $\utilityf(j, p) \ge \level_{ik}$. Since $j\not \in J$, and $p'_{j}=p_{j}$ for all $j\not\in J$. This yields $\utilityf(j, p') \ge \utilityf(h, p')$ for all $h\in \M_{k}(\A)\setminus J$, and $\utilityf(j, p') = \utilityf(j, p)\ge \level_{ik}$. Therefore, $j = A'_{ik}$ based on \cref{prop:priceLowerBound:eq:new-allo} and so $j\in \M(\A')$. This proves desired property $\M(\A') = \M(\A)\setminus J$, and furthermore, our argument implies $B_{j} \subseteq B'_{j}$ for $j \in \M_k(\A)$. Replacing $\M_{k}(\A)\setminus J$ by $\M_{k}(\A')$, 
\cref{prop:priceLowerBound:eq:new-allo} implies, if $A'_{ik}\neq \emptyset$ then
\[
\utilityf(A'_{ik}, p') \ge \utilityf(S, p'), \qquad S\subseteq \D_{i} \cap \M_{k}(\A')
\]
Otherwise, $A'_{ik}=\emptyset$ and $\utilityf(S,p') < \level_{ik}$. This is essentially the envy-free condition in \Cref{def:envy} and \Cref{lem:envy} claims $(\A, p)$ satisfies \cref{eq:envy}. We conclude that $(\A', p')$ is feasible.

Next we prove that the objective value of $(\A', p')$ is higher than $(\A, p)$. Indeed, 
\begin{align*}
\sum_{j\in \M(\A')}\left[(p'_{j}-c_{j})|B'_{j}| - d_{j} \right] &\ge \sum_{j\in \M(\A)\setminus J}\left[(p_{j}-c_{j})|B_{j}| - d_{j} \right] \\
&\ge \sum_{j\in \M(\A)}\left[(p_{j}-c_{j})|B_{j}| - d_{j} \right]
\end{align*}
where the first $\ge$ comes from that $B_{j} \subseteq B'_{j}$ and the second $\ge$ follows from the fact that $\sum_{j\in J}\left[(p_{j}-c_{j})|B_{j}| - d_{j} \right] < 0$ because $p_{j} < c_{j}$ for $j\in J$.
\end{proof}

\subsection{Chain equality on prices}

The next structural property we prove is that there exists optimal pricing satisfying a utility-chain equality. When the market is unit-supply and utility is linear, a similar result has been proven in \cite{chen2014envy}. We here provide a version in unlimited-supply market without linear utility assumption.

\begin{theorem}\label{utilitychain}
    For each optimal solution $(\A, p)$ to \cref{eq:prob-ep}, the following two conditions must be satisfied on each category set $M_k$.
    \begin{enumerate}
        \item The items with lowest marginal profit have prices equal to some buyers maximum willingness. For all $j \in \M^{\downarrow} := \arg\min_{j\in \M_k} (p_j-c_j)$,
        \[
        \exists i\in B_j, \quad  s.t.\quad  p_j=\R_{ij}
        \]
        \item For all other items $j\in \M_{k} \setminus \M^{\downarrow}$, there exists a chain equality
        \[
        \exists i \in B_j, \quad \exists j'\in \M_{k}\setminus\{j\}, \quad s.t. \quad \utilSing(j, p_j) = \max\{\level_{ik}, \ \utilSing(j', p_{j'})\}
        \]
        In particular, $p_{j'} - c_{j'} \le p_{j}- c_{j}$.
    \end{enumerate}    
\end{theorem}

To prove this theorem, we need a few technical results.

\begin{lemma}	\label{fixp}
Let $(\A, p)$ be any optimal solution  to \cref{eq:prob-ep}, and $j$ be any item with lowest profit margin in an item partition $\M_{k}$, i.e., $j\in \arg\min_{j\in \M_{k}}(p_j-c_j)$. Then there exists $i \in B_{j}$ such that $p_j = \R_{ij}$.
\end{lemma}
\begin{proof}
Without loss of generality, we assume there is just one item partition $\J$. By contradiction let for any optimal solution $(\A, p)$, suppose there exists $j^*\in \arg\min_{j\in \M}(p_j-c_j)$, but $p_{j^*}\neq \R_{ij^*}$ for all $i\in B_{j^*}$. Recall that the minimum utility level constraints \Cref{eq:prob:min-level} implies $p_j \le \min_{j\in B_j} \R_{ij}$.
We will provide a solution $(\A', p')$ with higher optimal value.  For brevity, we denote the items with lowest profit margin as $\M^{\downarrow} = \arg\min_{j\in \M}(p_j-c_j)$, and $\M':=\M(\A')$. The new price $p'$ is defined as
\begin{equation}\label{pj}
p'_j = \left\{ \begin{array}{ll}
p_j, &\forall \, j\not \in \M^{\downarrow}\\
\min\limits_{i\in B_j}\R_{ij}, &\forall \, j\in \M^{\downarrow}
\end{array}
\right.
\end{equation}
The construction of $p'$ guarantees $p'\ge p$ which implies $\utilSing(j, p'_j) \le \utilSing(j, p_j)$ for all buyer $i$ and item $j$. Therefore, any buyer $i$ unallocated in $\A$ remains unallocated under new price $p'$. In other words, $A'_i = A_i = \emptyset$ for all $i\not\in \B$. 
For buyers who have bought any item, they will never been unallocated because we only adjust prices for items in $\M^{\downarrow}$, and 
\[
\begin{aligned}
    &\utilSing(A_{i}, p'_{A_{i}}) = \utilSing(A_{i}, p_{A_{i}}) \ge \level_i, &\forall \, i\in \B \setminus \cup_{j\in \M^{\downarrow}} B_j \\
    &\utilSing(A_{i}, p'_{A_{i}}) \ge \utilSing(A_{i}, \R_{i, A_{i}}) \ge \level_i, &\forall \, i\in \cup_{j\in \M^{\downarrow}} B_j \\
\end{aligned}
\]
Therefore, there exists at least one item with utility no less than $\level_i$ for each buyer with $A_{i}\not = \emptyset$. Furthermore, we claim that $A_i \in \arg\max_{j\in \M} \utilSing(j, p'_j)$ for all $i\not \in \cup_{j\in \M^{\downarrow}} B_j$. Indeed, $\utilSing(A_i, p'_{A_i}) = \utilSing(A_i, p'_{A_i}) = \max_{j\in \M} \utilSing(j, p_j) \ge \max_{j\in \M'} \utilSing(j', p_j')$. Note that the last $\ge$ follows from that we only consider items within $\M$, and therefore $\M'\subseteq \M$. Under new price $p'$, only buyers in $\cup_{j\in \M^{\downarrow}} B_j$ may buy different things compared to $\A$. In summary, we define
\[
A'_i = \left\{
\begin{aligned}
&\in \arg\max_{j\in \M} \utilSing(j, p'_j), &i \in \cup_{j\in \M^{\downarrow}} B_j\\
 &   A_i, &\text{otherwise}
\end{aligned}
\right.
\] In case of multiple items maximising the buyers' utilities, we randomly choose one. The argument above shows the resulting $\A'$ satisfies minimum level requirement \cref{eq:prob:min-level} and also envy-free requirement \cref{eq:prob:envy}. All other requirements in \cref{eq:prob-ep} are also clearly satisfied. Consequently, $(\A', p')$ is feasible to \cref{eq:prob-ep}. To derive contradiction, it is sufficient to show $(\A', p')$ has higher objective value \cref{eq:prob:obj} than $(\A, p)$. 
\[
\begin{aligned}
 OBJ(\A', p') &= \sum_{j\in \M'} |B_j'|(p'_j-c_j) - \sum_{j\in \M'} d_j  \ge \sum_{j\in \M'} |B_j'|(p'_j-c_j) - \sum_{j\in \M} d_j \\
 &\ge \sum_{j\in \M'} |B_j'|(p_j-c_j) - \sum_{j\in \M} d_j 
\end{aligned}
\]
where $B'_j = B_j(\A')$ is the buyers who have bought item $j$ in $\A'$. Regarding the inequalities, the first $\ge$ comes from $\M'\subseteq \M$ and the second $\ge$ comes from $p'\ge p$. It follows that
\[
\begin{aligned}
  OBJ(\A', p') - OBJ(\A, p) &\ge \sum_{i\in \bigcup_{j\in \M^{\downarrow}} B_j} ((p'_{A'_{i}} - c_{A'_{i}}) - (p_{A_{i}} - c_{A_{i}}))  > 0
\end{aligned}
\]
where the first $\ge$ comes from that only buyers in $\cup_{j\in \M^{\downarrow}} B_j$ may change items to purchase and the second $>$ comes from the fact that $(p'_{A'_{i}} - c_{A'_{i}}) - (p_{A_{i}} - c_{A_{i}}) \ge 0$. This follows directly from $p'\ge p$ if $A'_{i}=A_{i}$. In the case $A'_{i}\neq A_{i}$, we know $p'_{A'_{i}} - c_{A'_{i}} \ge p_{A'_{i}} - c_{A'_{i}}\ge p_{A_{i}} - c_{A_{i}}$ from that $A_{i}\in \M^{\downarrow}$ for all $i\in \cup_{j\in \M^{\downarrow}} B_j$. In particular, we know $p'_{j^*} > p_{j^*}$ by construction of $p'$ and so we can assert $p'_{A'_{i}} - c_{A'_{i}} >  p_{j^*}-c_{j^*}$ for all $i\in B_{j^*}$. 
\end{proof}

The proof of this technical lemma is existential and not constructive. Hence, it does not tell us which $(i,j)$ pair the property holds for. However, it does imply an explicit formula for the easy case of having a single item. We only need to iterate over the pricing candidates $\{\R_{i1}: i\in \m \}$ to choose the one maximising the profit.

\begin{corollary} \label{cor:opt 1 item}
For $i \in \m$, let $n=1$ and set $S_i = \{ t \in \m : \R_{t1} \geq \R_{i1}\}$.  Then the optimal solution is $\mu^* = \left[\max_{i \in \m} \{ (\R_{i1} - c_1) |S_i| \} - f_1\right]^+ $,  with $p_1^*= \R_{i^*1}$ if $\mu^* >0$ and equal to 0 if $\mu^* = 0$, where $i^* = \arg\max_{i \in \m} \{ (\R_{i1} - c_1) |S_i| \} - f_1$. 
\end{corollary}

The idea of \Cref{fixp} is to increase the price of these items with minimum marginal profit $p_j-c_j$. The same idea leads to a chain rule on the utility for optimal pricing.

\subsection{Proof of \Cref{utilitychain}}

Without loss of generality, we drop the category index $k$ (e.g. $\M$ instead of $\M_k$) throughout this proof. 
    The first condition is proved by \Cref{fixp}. For the second condition, by contradiction suppose there exists an optimal allocation $(\A, p)$ such that $\exists j^*\in \M \setminus \M^{\downarrow}$ such that $\utilSing(j^*, p_{j^*}) \neq \utilSing(j', p_{j'})$ for all $i\in B_{j^*}$ and $j'\in \M^{\le}_{j^*}:=\{j\in \M: p_{j}-c_{j} \le p_{j^*}-c_{j^*}\}$. Since $i\in B_{j^*}$, from envy-freeness \cref{eq:prob:envy}, we know $\utilSing(j^*, p_{j^*}) \ge \utilSing(j,p_{j})$ for all $j\in \M$ and also minimum level \cref{eq:prob:min-level} requires $\utilSing(j^*, p_{j^*}) \ge \level_{i}$. Therefore, it is equivalent to say
    \begin{equation}\label{eq:neqzero}
        \delta(p):=\min_{i\in B_{j^*}}\min_{j\in \M^{\le}_{j^*}} \left[\utilSing(j^*, p_{j^*}) - \max\{\level_{i}, \ \utilSing(j, p_{j})\}\right] > 0
    \end{equation}
    We aim to construction a new solution $(\A', p')$ with higher objective value, and so obtain a contradiction to the optimality of $(\A, p)$. The new price $p'$ is defined as same as $p$ except for item $j^*$, we set $p_{j^*}$ to be the root of $\delta(p)$. Clearly, we have $p'_{j^*} > p_{j^*}$. 
    For the new allocation $\A'$, we will enforce $\M':=\M(\A') \subseteq \M$. This is legit by  delisting all items in $M\setminus \M$ when making allocation $\A'$. Then $\A'$ is determined depending on the buyers. First, for buyer $i\not \in B_{j^*}$,
    we set $A'_{i}=A_{i}$. This is valid because $p'\ge p$ and $p'_{A_{i}}=p_{A_{i}}$ when $A_{i}\neq j^*$. More precisely, the utility of $A_{i}$ is still dominant under price $p'$: for all $j\in\M$,
    \[
    \utilSing(A'_{i}, p'_{A'_{i}}) = \utilSing(A_{i}, p'_{A_{i}}) = \utilSing(A_{i}, p_{A_{i}}) \ge \max\{\level_{i}, \utilSing(j, p_{j})\} \ge \max\{\level_{i}, \utilSing(j, p'_{j})\}.
    \]
For buyer $i\in B_{j^*}$, however, we set $A'_{i} \in \arg\max_{j\in \M} \utilSing(j, p'_j)$ by picking item maximising their utility. If there are multiple such items, we pick any one with the highest marginal profit. 
    The envy-freeness follows directly from this construction. It remains to verify the minimum level constraint $\utilSing(A'_{i}, p'_{A'_{i}}) \ge \level_{i}$. Recall that $p'_{j^*}$ is the root of \cref{eq:neqzero} and $i\in B_{j^*}$. It follows that
    \[
    \utilSing(A'_{i}, p'_{A'_{i}}) \ge \utilSing(j^*, p'_{j^*}) \ge \level_{i}
    \]
    Consequently, $(\A', p')$ is feasible to \cref{eq:prob-ep}. Similar to the proof of \Cref{fixp}, let $B'_j=B_j(\A')$:
    \[
    \begin{aligned}
        OBJ(\A', p') &= \sum_{j\in \M'} |B_j'|(p'_j-c_j) - \sum_{j\in \M'} d_j  \ge \sum_{j\in \M'} |B_j'|(p'_j-c_j) - \sum_{j\in \M} d_j \\
 &\ge \sum_{j\in \M'} |B_j'|(p_j-c_j) - \sum_{j\in \M} d_j 
    \end{aligned}
    \]
    Recall that $A'_{i}=A_{i}$ for all $i\not\in B_{j^*}$. 
    \[
    \begin{aligned}
        OBJ(\A', p') - OBJ(\A, p) \ge \sum_{i\in B_{j^*}} ((p'_{A'_{i}}-c_{A'_{i}})-(p_{j^*}-c_{j^*}))
    \end{aligned}
    \]
    If $A'_{i}=j^*$, then we have $p'_{A'_{i}} > p_{j^*}$. Otherwise, $A'_{i}\neq j^*$. We claim $p'_{A'_{i}}-c_{A'_{i}} > p_{j^*}-c_{j^*}$. Indeed, \Cref{fixp} guarantees item $j^*$ has at least the same utility as all items in $\M^{\le}_{j^*}$. Since $A'_{i}\neq j^*$, then we must have $p'_{A'_{i}}-c_{A'_{i}} \ge p'_{j^*}-c_{j^*} > p_{j^*}-c_{j^*}$. It follows that $OBJ(\A', p') > OBJ(\A, p)$.

\subsection{Integrality under Montonicity}

When the utility functions are order-preserving under rounding up of prices, we argue that the optimal prices are integer valued. This integrality property of prices is not only interesting on its own but is also useful for complexity analysis in the next section. The conditions required to apply this result are not overly restrictive. For example, linear utility $\utility = \valuation - p_j$ with $\valuation\in \mathbb{Z}$ satisfies them.

\begin{proposition}\label{intPrice}
Suppose buyers are \textit{unit-demand}, the maximum willingness $\R_{ij}$ are all integers, and 
$\utilityf$ satisfies
\begin{equation}\label{eq:intPriceCondUtility}
    \utilityf(i, p) \ge \utilityf(i', p) \quad \implies \quad \utilityf(i, \ceil{p}) \ge \utilityf(i', \ceil{p})
\end{equation}
where $\ceil{p}$ is the price after rounding up, i.e. $\ceil{p}_i = \ceil{p_i}$ for all $i$. Then, for every optimal price $p^{\ast}$, the total price on a buyer's allocation in every category is integral, i.e. $p^{\ast}_{A_{ik}} \in \mathbb{Z}$ for all $i\in \m, k\in [r]$.
\end{proposition}	
\begin{proof}
For sake of contradiction, let for any optimal solution $(\A, p)$ there exist some $i\in \m$ such that $p_{A_{i}}\not \in  \mathbb{Z}$. Now we define a new price $\bar{p}$ such that $\bar{p} = \ceil{p}$.
We will prove $(\A, \bar{p})$ is a feasible solution with higher objective value, leading to a contradiction that $(\A, p)$ is optimal. Since we only change the price, it is sufficient to verify constraints on the minimum level of utilities \cref{eq:prob:min-level} and envy-free constraints \cref{eq:prob:envy}. Take any buyer $i$ and item category $k$. 

Since $\utilityf(A_{ik}, p) \ge \level_{ik}$ and $\utilityf(A_{ik}, \R_{i, A_{ik}}) = \level_{ik}$, the monotonicity of $\utilityf$ implies $p \le \R_{i, A_{ik}}$. By our assumption, $\R_{i, A_{ik}}\in \mathbb{Z}$ means $p'_{A_{ik}}=\ceil{p_{A_{ik}}} \le \R_{i, A_{ik}}$. Therefore, $\utilityf(A_{ik}, p'_{A_{ik}}) \ge \utilityf(A_{ik}, \R_{i, A_{ik}}) = \level_{ik}$, which verifies \cref{eq:prob:min-level}.

Observe that, for all $S\in \D_{i}\cap \M(\A) \cap M_{k}$. If $\level_{ik} \ge \utilityf(S, p_{S})$, then $\level_{ik}\ge \utilityf(S, \ceil{p_{S}})$ due to $\utilityf(S, p_{S}) \ge \utilityf(S, \ceil{p_{S}})$. For other sets $S$, if $\utilityf(A_{ik}, p_{A_{ik}}) \ge \utilityf(S, p_{S})$, then \cref{eq:intPriceCondUtility} guarantees $\utilityf(A_{ik}, \ceil{p_{A_{ik}}}) \ge \utilityf(S, \ceil{p_{S}})$, thus verifying \cref{eq:prob:envy}.


It is obvious that the objective of $(\A, \bar{p})$ is higher than $(\A, p)$ due to $\bar{p}\ge p$. This contradicts the optimality of $(\A, p)$.
\end{proof}

\section{Complexity Analysis}\label{sec:complexity}

\citet{guruswami2005profit} established the APX-hardness of the problem when each buyer can purchase multi-items as a bundle. Such hardness remains even when the market has unlimited supply for each item, and each buyer is \textit{single-minded} for a bundle with cardinality being 2. In contrast, when each buyer is forced to buy bundles with cardinality being 1, this is so-called \textit{unit-demand}. \citet{chen2014envy} examined \textit{unit-demand} and \textit{unit-supply} scenarios where each item has only 1 copy available. They proved the problem is NP-hard when each buyer prefers at least three items, but polynomial solvable if all buyers prefer at most two items.

Our study extends these findings to situations with \textit{unlimited supply} and \textit{unit-demand} buyers. We find that the envy-free pricing problem is polynomially solvable when each buyer is interested in only one specific item. Also, the problem becomes NP-hard  when each buyer desires at least two items. This  supplements the previous single-minded scenario.

\begin{table}[t!]
\small
    \centering
    \begin{tabular}{l|l|l|l}
    \toprule
        Unit-demand & Unlimited-supply & Items/bundles desired & NP-hard \\ \midrule
         No & Yes & $1$ & Yes \cite{guruswami2005profit} \\\midrule
         Yes & No & $\le 2$ &  No \cite{chen2014envy} \\\midrule
         Yes & No & $3$ &  Yes \cite{chen2014envy} \\\midrule
         Yes & Yes & $1$ & No, \Cref{thm:poly-solvable} \\\midrule
         Yes & Yes & $\le 2$ & Yes \cite{guruswami2005profit} \\\midrule
         Yes & Yes & $2$ & Yes, \Cref{thm:np-hard} \\
         \bottomrule
    \end{tabular}
    \caption{Complexity of envy-free pricing problems \cref{eq:prob-ep}}
    \label{tab:complexity}
\end{table}

Additionally, we prove the envy-free pricing problem is pseudo-polynomially solvable for a category of utility functions including linear utility in \Cref{thm:pseudo-poly}. These results highlight the impact of buyer preferences on the complexity of achieving envy-free pricing, providing new insights into the conditions under which market fairness is computationally feasible. 

\begin{proposition}\label{thm:poly-solvable}
Suppose the market is \textit{unit-demand} and \textit{unlimited-supply}, and each buyer is \textit{single-minded}. Then, an envy-free pricing  is polynomially solvable.
\end{proposition}
\begin{proof}
    Let $ r_i $ denote the single item preferred by buyer $ i $, where $ r_i $ is defined such that $ R_{ir_i} > 0 $. Define $ I_j = \{i \in \m : r_i = j\} $ as the set of buyers preferring item $ j $. Since each buyer $ i $ is single-minded, $ R_{ij} = 0 $ for all $ j \neq r_i $, ensuring buyer $ i $ will only choose item $ r_i $ regardless of others' price.
    
    Consider any optimal pricing scheme $(\A, p)$. If $i\in B_j$, then $j=r_{i}$. 
    The utility condition \Cref{utilitychain} implies $p_j=\min_{i\in B_j} \R_{ij}$ unless there exists another item $j'\in \n$ with equation $\utilSing(j, p_j) = \utilSing(j', p_{j'})$ for some $i \in B_{j}$. Since $j'\neq j = r_{j}$, $\utilSing(j', p_{j'}) \le \utilSing(j', 0) = 0$ which implies $\utilSing(j, p_j) = 0$ i.e. $p_j = \R_{ij}$.
    
    Therefore, if $ B_j $ is non-empty, then $ p_j = \min_{i \in B_j} \R_{ij} $. Since the price $ p_j $ affects only the choices of buyers in $ I_j $, the optimal price for item $ j $ can be computed in polynomial time applying the formula \Cref{cor:opt 1 item} to the sub-market with $ I_j $ and item $ j $ alone. Repeating this for each item results in an optimal pricing scheme $ (\A, p) $ in polynomial time.
\end{proof}

A natural question to ask is to which extent the \textit{unit-demand} and \textit{unlimited-supply} envy-free pricing problem shifts from polynomial solvable to become NP-hard. We explore such transition in terms of number of items each buyer desires. The following theorem proves the problem is NP-hard even when each buyer desires just 2 items.

\begin{theorem}\label{thm:np-hard}
    The \textit{unit-demand} envy-free pricing with \textit{unlimited-supply} is NP-hard even if each buyer desires at least 2 items. 
\end{theorem}
\begin{proof}
We reduce from maximum 2-SAT problem (MAX 2-SAT) which is known to be both NP-hard and APX-hard \cite{haastad2001some}. This problem seeks to determine the maximum number of satisfiable 2-clauses of the form $(x \vee y)$ can be true, where $x, y$ are boolean variables that can each be either true or false. Each boolean variable $x$ also has its negation $\bar{x}$ appearing in some clauses. 
    
    For the MAX 2-SAT problem with $k$ clauses denoted by $\mathbf{cl}_1, \ldots, \mathbf{cl}_k$, and $2h$ distinct boolean variables denoted by $x_1, \bar{x}_1, x_2, \ldots, x_h, \bar{x}_h$. For each boolean variable, we create an item. Specifically, for each variable variables $x_{i}$ and its negation $\bar{x}_{i}$, items $j_{i}$, $\bar{j}_{i}$ are created respectively. Then, we introduce $12hk+k$ buyers divided into four types. 
    \begin{enumerate}
        \item \textbf{First type:} There are $k$ buyers $\alpha_1,\ldots, \alpha_{k}$, where each buyer desires the two items corresponding to the variables in clause $\mathbf{cl}_i$ with a valuation of 1.
        \item \textbf{Second type:} Consisting of $8kh$ buyers distributed evenly among $h$ pairs of items $j_i$ and $\bar{j}_i$. Each pair attracts $8k$ buyers $\beta_1(i),\ldots \beta_{8k}(i)$, each with a valuation of $1$.
        \item \textbf{Third type:} This group includes $2kh$ buyers, where each of the $h$ item pairs $j_i$ and $\bar{j}_i$ attracts $2k$ buyers $\gamma_{1}(i), \ldots, \gamma_{2k}(i)$. The valuation for each buyer is $1$ for $j_i$ and $3$ for $\bar{j}_i$.
        \item \textbf{Fourth type:} Similarly, there are $2kh$ buyers, with $2k$ buyers $\delta_{1}(i), \ldots, \delta_{2k}(i)$ per item pair $j_i$ and $\bar{j}_i$. Each buyer values $j_i$ at $3$ and $\bar{j}_i$ at $1$.
    \end{enumerate}
    In a summary, there are $2h$ items: $\n=\{j_1, \bar{j}_1, j_2, \bar{j}_2,\ldots, j_h, \bar{j}_h\}$, and the $12hk+k$ buyers: 
    $$
    \m = \left\{\begin{aligned}
        &\alpha_1,\ldots, \alpha_k,\\
        &\beta_1(1),\ldots,\beta_{8k}(1), \beta_1(2),\ldots,\beta_{8k}(2), \ldots, \beta_1(h),\ldots, \beta_{8k}(h), \\ 
        &\gamma_{1}(1),\ldots,\gamma_{2k}(1), \gamma_{1}(2),\ldots,\gamma_{2k}(2), \ldots, \gamma_1(h),\ldots, \gamma_{2k}(h), \\
        &\delta_{1}(1),\ldots,\delta_{2k}(1),\delta_{1}(2),\ldots,\delta_{2k}(2), \ldots, \delta_1(h),\ldots, \delta_{2k}(h),
    \end{aligned}\right\}
    $$
    The utility of buyers are all linear $\utilSing(j, p_j) = \valuef(j)-p_j$ and the minimum utility levels are all $1$. We assume the cost of items and their listing fee are all $0$. In other words, the seller's profit is simply the total sales $\sum_{j\in \M} p_j |B_j|$. Next, we derive the equivalence between MAX 2-SAT and constructed envy-free pricing problem. Specifically, we prove the following two properties: 
    \begin{enumerate}
        \item If MAX 2-SAT has $t$ clauses satisfiable, then the envy-free pricing problem's optimal objective is at least $t+16kh$.
        \item Conversely, if envy-free pricing has optimal objective being $t+16kh$, then MAX 2-SAT has at least $t$ clauses satisfiable. 
    \end{enumerate}
    The first claim is straightforward. Suppose MAX 2-SAT has $t$ clauses satisfiable with true-false assignment $x_1, \bar{x}_1,\ldots, x_h, \bar{x}_h$. For each boolean variable pair $x_i, \bar{x}_i$, if $x_i$ is true, then prices item $j_i$ at 1 and $\bar{j}_i$ at 3, otherwise prices item $j_i$ at 3 and $\bar{j}_i$ at 1. Under such pricing, there exists a allocation $\A$ with profit being $t+16kh$. Indeed, buyer $\alpha_i$ purchases an item with price $1$ if clause $\mathbf{cl}_i$ is true, otherwise $\alpha_i$ purchase nothing because both items priced at $3$ which is larger than $\alpha_i$'s valuation which is $1$. Therefore, all first type buyers contribute $t$ to the revenue.
    Each buyer $\beta_l(i)$ contributes exactly 1 to the revenue because there is always an item between $j_i$ and $\bar{j}_i$ being priced at 1, resulting into $8kh$ total revenue from second type buyers.
    For the third type and forth type buyers, observe that the total revenue from $\gamma_l(i)$ and $\delta_l(i)$ is exactly $4$, since exactly one between $j_i$ and $\bar{j}_1$ has price 1 and the other has price $3$. These two items are chosen separated by $\gamma_l(i)$ and $\delta_l(i)$. Consequently, the combined revenue contributions from third and fourth type buyers are $8kh$. Therefore, the total revenue of the seller is $t+16kh$. 

    For the reverse direction, suppose the optimal solution to envy-free pricing is $(\A, p)$ with objective value being $t+16kh$. It's sufficient to show that, for each item pair $j_i$ and $\bar{j}_i$, exactly one is priced at 1 and the other priced at 3. Then we convert price $p$ into true assignment to the MAX 2-SAT by setting true to all  boolean variables corresponding to item with price 1, and false to others.
    
    Since the utility is linear and all valuations are integers, \Cref{intPrice} implies optimal prices are integral. Let  $p \in \mathbb{Z}^{2h}$ be an optimal price. \Cref{lem:boundprice,prop:priceLowerBound} imply $0\le p\le 3$. 
    Observe that there are $k+8hk$ buyers which is less than the revenue $t+16kh$. There must be some items with prices larger than $1$. However, not all items can have price larger than 1. More precisely, for each item pair $j_i$ and $\bar{j}_i$, at least one between them must be priced at $1$ and the other has price larger than 1. We classify all item pairs into three sets $I_1$, $I_2$, and $I'$: $I_1=\{i\in [h]: \min\{p_{j_i},p_{\bar{j}_i}\} \le 1, \max\{p_{j_i},p_{\bar{j}_i}\} = 1\}$, $I_2=\{i\in [h]: \min\{p_{j_i},p_{\bar{j}_i}\}=1, \max\{p_{j_i},p_{\bar{j}_i}\} \ge 2\}$ and $I' = [h]\setminus (I_1\cup I_2)$. Note that $I'$ means the set of item pairs neither have prices being $1$.  
    If $I'\cup I_1\not\eq \emptyset$, then the total revenue must be smaller than $t+16kh$. Indeed, for each $i\in I'$, revenue contributions from buyers $\beta_l(i)$ towards items pair $j_i$ and $\bar{j}_{i}$ is $0$ due to the valuations being $1$ but the prices are either $0$ or larger than $1$ resulting into no revenue or unable to purchase respectively. For $\gamma_l(i)$ and $\delta_l(i)$, the maximum possible revenues are $6k$ towards these two items. Therefore, the total revenue from buyer types in second, third, and fourth is at most $12k$ for each $i\in I'$. For $i\in I_1$, the total revenue is also at most $12k$ from $12k$ buyers in last three types because both items in a pair $i$ are priced at most $1$.   
    For $i\in I_2$, however, the last three type of buyers may contribute $16k$ to the total revenue. This is obtained by simply adding all the maximum valuations from these buyers. Consequently, the maximum total revenues are $12k(|I'|+|I_1|) + 16k|I| + k$ where the last $k$ is an upper bound for the revenue contributed by $k$ buyers in the first type. Since $|I'|+|I_1| + |I_2|=h$, we have objective revenue is $t+16kh \ge 16kh$ which is larger than $16kh + k - 4k(|I'|+|I_1|)$ if $|I'| + |I_1| > 0$. Therefore, we conclude that $I'=I_1=\emptyset$, i.e. each item pair $j_i$ and $\bar{j}_i$ must have at least one priced at $1$ and the other has price at least 2. 
    Moreover, we know the price of the more expensive item in pair $i$ must be $3$. This is because such item can only be chosen by buyers in last two types with valuation at $3$ towards it and these buyers having zero utility against all other items, so \cref{utilitychain} implies the utility for the chosen item is also $0$ meaning the price must be $3$. 
Thus, for each item pair $j_i$ and $\bar{j}_i$, there is one priced at 1 and the other priced at 3. 

It is straightforward to calculate the revenue from buyers in last three types is $16kh$ in total. To make up the total revenue $t+16kh$, the buyers  in first type must contribute exactly $t$. Recall that buyer $\alpha_l$ desires two items corresponding to clause $\mathbf{cl}_l$ with valuation being $1$. If $\alpha_l$ purchases any item, then such item must be priced at $1$, which means the corresponding boolean variable is true and so is the clause $\mathbf{cl}_l$. Consequently, after assigning true to items priced at 1 and false to others, we obtain a true assignment for MAX 2-SAT with $t$ clauses satisfied. 

We have argued the equivalence between solutions of MAX 2-SAT and envy-free pricing. The construction of envy-free pricing is clearly in polynomial time and therefore, this is a polynomial-time reduction. Consequently, the unit-demand and unlimited-supply envy-free pricing with buyers desiring at least 2 items is both NP-hard and APX-hard.
\end{proof}

If we further impose monotonicity condition \cref{eq:intPriceCondUtility} on the utility function, then we achieve pseudo-polynomial solvability.

\begin{theorem}\label{thm:pseudo-poly} 
For \textit{unit-demand} market, if the utility function satisfies \cref{eq:intPriceCondUtility}, and maximum willingness $\R_{ij}$ are integer valued for all $i,j$, then problem~\cref{eq:prob-ep} is solvable in 
pseudo-polynomial time for bounded $n$.
\end{theorem}

\begin{proof}
\Cref{intPrice} tells us that the optimal pricing must be integral valued. Since $p_j\le \max_{i\in \m} \R_{ij} \le \norm{\R}_{\infty}$ from \Cref{lem:boundprice} for each $j\in \n$, there are at most $\norm{\R}_{\infty}^{n}$ candidates for $p_1,p_2,\ldots, p_n$.  Once we fix price $p':=(p'_1,p'_2,\ldots, p'_n)$ and which items to sell, then a profit-maximising feasible allocation can be found through a weighted matching problem on a bipartite graph. There are $2^n$ candidates for which items to sell. For $p$, we only need to consider integer values by \Cref{intPrice}. Therefore, we have $(2\norm{A}_{\infty})^n$ candidates to fix and for each of them it takes polynomial time for the matching problem. Thus, pseudo-polynomial solvability follows for bounded $n$.
\end{proof}

\paragraph{Acknowledgements}
AI was not used in deriving the research ideas or writing any parts of this paper. The authors have no competing interests to declare that are relevant to the content of this article.

{
\newrefcontext[sorting=nyt]
\printbibliography
}

\end{document}